\documentclass[prl,superscriptaddress,showpacs, twocolumn,floatfix,amsmath,footinbib,amssymb]{revtex4-1}
\usepackage{amssymb}
\usepackage{amsmath}
\usepackage{comment}
\usepackage{epsfig}
\usepackage{t1enc}
\usepackage{soul}
\usepackage{color}
\newcommand{\beq}{\begin{equation}}
\newcommand{\eeq}{\end{equation}}
\newcommand{\bea}{\begin{eqnarray}}
\newcommand{\eea}{\end{eqnarray}}

\begin{document}

\title{Entanglement and discord: accelerated observations of local and global modes.}
\author{Jason Doukas}
\address{National Institute for Informatics, 2-1-2 Hitotsubashi, Chiyoda-ku, Tokyo 101-8430,  Japan}
\author{Eric G. Brown}
\address{Department of Physics and Astronomy, University of Waterloo, Waterloo, Ontario N2L 3G1, Canada}
\author{Andrzej Dragan}
\address{Institute of Theoretical Physics, University of Warsaw, Ho\.{z}a 69, 00-049 Warsaw, Poland}
\author{Robert B. Mann}
\address{Department of Physics and Astronomy, University of Waterloo, Waterloo, Ontario N2L 3G1, Canada}
\address{ Perimeter Institute for Theoretical Physics, Waterloo, Ontario N2L 2Y5, Canada}

\date{\today}
\begin{abstract}
We investigate the amount of entanglement and quantum discord extractable from a two mode squeezed state as considered from the viewpoint of two observers, Alice (inertial) and Rob (accelerated). We find that using localized modes produces qualitatively different correlation properties for large accelerations than do Unruh modes. Specifically, the entanglement undergoes a sudden death as a function of acceleration and the discord asymptotes to zero in the limit of infinite acceleration. 
We conclude that the previous Unruh mode analyses do not determine the acceleration dependent entanglement and discord degradation of a given quantum state.
\end{abstract}
\maketitle

\emph{--Introduction.\thinspace}
The quest to understand the nature of entanglement in accelerated frames has recently taken a new turn. After it was realized that the entanglement degradation found in previous studies \cite{Bruschi2010} of Unruh mode entanglement in non-inertial frames had to be reinterpreted, a new programme to understand this problem from a localized perspective was initiated \cite{Dragan2012, Dragan2012b}. Our current work bridges results recently found with localized Gaussian modes with those  previously obtained using Unruh modes. 

While previous studies have been mainly focussed on investigating the effects of acceleration on initially entangled Fock states \cite{Alsing2003,Fuentes2005,Datta2009, Bruschi2010,Brown2012}, calculations involving localized modes quickly become challenging in this setup and there is some evidence that these problems are computationally difficult \cite{Montero2012}. Using two-mode squeezed states instead allows us to do much more \cite{Dragan2012b}.   
While these states have been previously considered in the context of entanglement in non-inertial frames under the single mode approximation \cite{Adesso2007},  such analysis is not transferrable to  localized modes \cite{Dragan2012b}.  Since closed form expressions for  logarthmic negativity \cite{Adesso2004} and (Gaussian) quantum discord \cite{Adesso2010} can be calculated from the parameters of the covariance matrix, it is opportune to take advantage of the analytic expression for the covariance matrix calculated in \cite{Dragan2012b} to gain further insight into the nature of quantum information in accelerated frames.

Discord \cite{Ollivier2001} is a generalized measure of quantum correlations that, for a mixed state, can be nonzero even if the state is separable. It is computed by an optimization procedure over all possible measurements that can be performed on one of the subsystems, and in the Gaussian case \cite{Adesso2010} the optimization is restricted to Gaussian measurements only. Although entanglement is often the primary resource for many quantum computational procedures, it is now understood that quantum discord can also take operational significance and can be used as a resource for mixed-state quantum computing, even in the absence of entanglement \cite{operationaldiscord}. Therefore understanding the robustness of both entanglement and discord in non-inertial frames is an important goal of relativistic quantum information theory. A common theme in the literature on discord is that it appears to be more robust in the face of decohering noise than is entanglement, and previous studies of discord in non-inertial frames \cite{Datta2009,Brown2012} have shown that its degradation due to the contribution from the vacuum excitations is weaker than or comparable to that of entanglement, depending on the specific measure being considered.

In this paper we look at the maximal amount of quantum correlations that can be extracted from a fixed quantum state observed by two observers, one of whom is accelerating. We use this optimum measurement strategy to study differences between localized \cite{Dragan2012b} and unlocalized \cite{Alsing2003,Fuentes2005,Datta2009,Brown2012} modes in the correlations measured. In comparing localized Gaussian modes and global Unruh modes, we find two interesting differences. In the localized case the entanglement undergoes sudden death as a function of acceleration which was unknown in the previous literature. Secondly, for localized modes the discord vanishes for both Alice and Bob in the limit of infinite acceleration, a result which was also absent in the previous literature, see for example \cite{Datta2009}. These differences suggest that previous results from global mode analyses should not be expected to be seen in real experiments on acceleration-dependent entanglement degradation of a given state. \


We also consider the quantum correlations of Unruh modes as a function of Unruh frequency. It was shown \cite{Datta2009} for entangled Fock states in the limit of zero Unruh frequency \cite{Alsing2003,Fuentes2005}, that while the state is separable the quantum discord, as computed by optimizing over measurements on Alice's subsystem, does not vanish. We find the same behaviour for a two-mode squeezed state. We then go beyond this and find that the discord when the measurements are optimized over Rob's subsystem in fact vanishes. Hence there is a fundamental asymmetry to quantum discord in non-inertial frames, a fact hitherto unrealized because of the computational challenges involved in optimizing the measurement over Rob's subsystem in the entangled Fock state setting.

\emph{--Set up.\thinspace}
Consider an arbitrary state of the field of two modes $\hat{a}$ and $\hat{b}$, where $\hat{a}$ ($\hat{b}$) is the annihilation operator associated with Alice's (Bob's) mode defined by the classical, possibly unlocalized, wave-packet $\phi_A$ ($\phi_B$). To form valid annihilation operators (upon quantization of the field) we assume that $\phi_A$ and $\phi_B$ are superpositions of positive frequency Minkowski plane waves only.

\begin{figure}[h]
       \centering
        \includegraphics[width=0.5\textwidth]{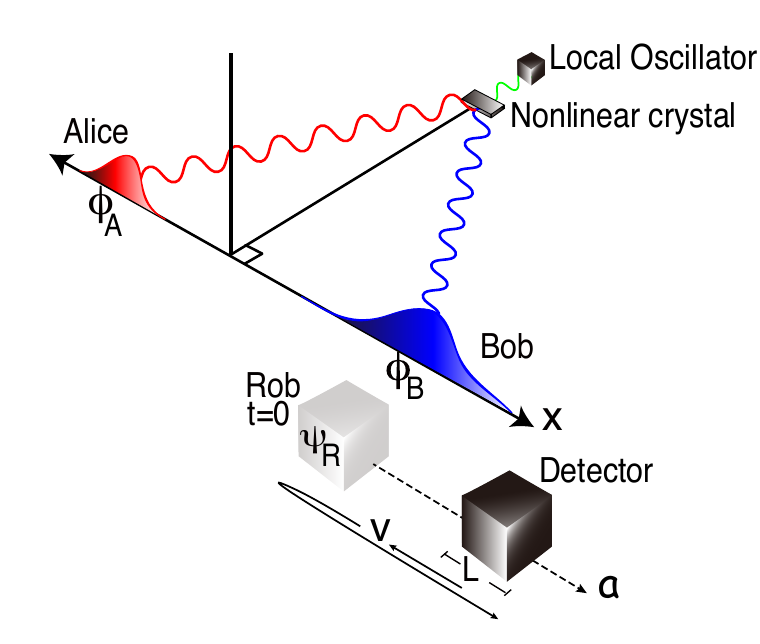}
	\caption{(Color
  online) A schematic of the type of setup we consider. A two-mode squeezed state is produced from a non-linear crystal in two Gaussian modes $\phi_A$ (Alice) and $\phi_B$ (Bob). An observer Rob accelerating with constant proper acceleration $a$ in the $x$-direction carries a detector that makes measurments of the field in a mode $\psi_R$ at time $t=0$ when his velocity, $v$, is zero.}
        \label{fig:scheme}
\end{figure}

Our intention is to view this state from the perspective of two observers: Alice who is inertial and has access to mode $\phi_A$, and Rob, who is uniformly accelerated but who only has (partial) access to Bob's part of the state $\phi_B$, see Fig.~\ref{fig:scheme}. Under accelerated motion the Minkowski vacuum state undergoes a well-known squeezing transformation \cite{Unruh1976} which in general mixes up Alice's and Bob's modes. To ensure that the  accelerated observer can distinguish Bob's subsystem from Alice's we further assume that $\phi_A$ and $\phi_B$ are distinguished by their parity taking Alice's (Bob's) mode to be a left (right) mover.

For the Bell-type entangled Fock state commonly used in previous analyses \cite{Alsing2003,Bruschi2010,Datta2009} the considered state is given by $(1+\hat{a}^{\dagger}\otimes \hat{b}^{\dagger})|0\rangle_M$, however, here we are primarily interested in the two-mode squeezed state which takes the form $\exp{\left[s(\hat{a}^{\dagger}\hat{b}^{\dagger}+\hat{a}\hat{b})\right]}|0\rangle_M$, with squeezing parameter $s$. Recall that a (zero mean) Gaussian state is fully characterized by its second quadrature moments, given by the covariance matrix \(\sigma_{ij}\equiv \langle \hat{X}_i \hat{X}_j +\hat{X}_j \hat{X}_i  \rangle\), where for a two-mode system \(\hat{\boldsymbol{X}}=(\hat{x}_A,\hat{p}_A,\hat{x}_B,\hat{p}_B)\) and the quadrature operators of a given mode are related to the annihilation and creation operators of that mode by \(\hat{x}=(\hat{a}+\hat{a}^\dagger)/\sqrt{2}\) and \(\hat{p}=-i(\hat{a}-\hat{a}^\dagger)/\sqrt{2}\). 

We suppose Alice (Rob) is in the possession of an inertial (accelerating) detector that couples to a mode of the field $\psi_A(x,t)$ ($\psi_R(\xi,\tau)$). We use the family of conformal Rindler coordinates $(\xi,\tau)$ parameterized by $a$ (in units $c=1$): 
\begin{equation}
t = a^{-1}e^{a\xi} \sinh a\tau, ~~~x = a^{-1}e^{a\xi} \cosh a\tau,
\end{equation}
to cover the region of Minkowski space accessible to Rob and assume that he moves along the line $\xi=0$ so that $a$ corresponds to his proper acceleration and $\tau$ to his proper time. The covariance matrix of the state $\exp{\left[s(\hat{a}^{\dagger}\hat{b}^{\dagger}+\hat{a}\hat{b})\right]}|0\rangle_M$ as observed by Alice and Rob was calculated in \cite{Dragan2012b} and is given by:
\begin{widetext}
\begin{eqnarray}
\label{covariance}
\sigma &=&
\openone+2\langle\hat{n}_U\rangle
\begin{pmatrix}
0&0&0&0\\
0&0&0&0\\
0&0&1&0 \\
0&0&0&1
\end{pmatrix}+2\sinh^2 s
\begin{pmatrix}
|\alpha|^2&0&0&0\\
0&|\alpha|^2&0&0\\
0&0&|\beta+\beta'^\star|^2&2\,\text{Im}(\beta\beta') \\
0&0&2\,\text{Im}(\beta\beta')&|\beta-\beta'^\star|^2
\end{pmatrix}\nonumber \\ \nonumber \\ 
& &+ 
\sinh 2s\begin{pmatrix}
0&0&-\text{Re}[\alpha(\beta+\beta'^\star)]&-\text{Im}[\alpha(\beta-\beta'^\star)]\\
0&0&-\text{Im}[\alpha(\beta+\beta'^\star)]&\text{Re}[\alpha(\beta-\beta'^\star)]\\
-\text{Re}[\alpha(\beta+\beta'^\star)]&-\text{Im}[\alpha(\beta+\beta'^\star)]&0&0 \\
-\text{Im}[\alpha(\beta-\beta'^\star)]&\text{Re}[\alpha(\beta-\beta'^\star)]&0&0
\end{pmatrix},
\end{eqnarray}
\end{widetext}
where $\alpha\equiv(\psi_A,\phi_A)$, $\beta\equiv(\psi_R,\phi_B)$, $\beta'\equiv(\psi_R,\phi^\star_B)$, the bracket is the usual Klein-Gordon scalar product, $(f,g)=i\int f^*\overleftrightarrow{\partial}_t g|_{t=0}dx$, and:
\begin{equation}
\label{noise}
\langle\hat{n}_U\rangle =  \int \frac{|(\psi_\text{R},w_{Ik})|^2}{e^{\frac{2\pi |k| }{a}}-1}  dk,
\end{equation}
is the average particle number Rob would measure if he were accelerating through the Minkowski vacuum. $w_{Ik}=\frac{1}{\sqrt{4\pi |k| }}e^{i(k\xi-|k|\tau)}$ is the region I positive frequency Rindler mode.

We assume that Alice's detector $\psi_A$ perfectly detects the mode $\phi_A$, i.e., $\psi_A=\phi_A$, and for each acceleration Rob chooses a detector that maximizes the entanglement extracted from the considered state. It has been shown in \cite{Dragan2012b} that the optimization over Rob's detector is limited by the amount of the state that Rob can observe due to the presence of the event horizon. Rob's optimal mode for a given acceleration is given by \cite{Dragan2012b}:
\bea\label{RobOptDet}
\psi_R=\frac{\int_{\Lambda}^{\infty}dk(w_{Ik},\phi_B)w_{Ik}}{\sqrt{\int_{\Lambda}^{\infty}dk|(w_{Ik},\phi_B)|^2}},
\eea
where $\Lambda$ is the cut-off frequency of the detector \cite{footnoteCutoff}

For this choice of Alice's and Rob's detectors $\alpha=1$ and:
\bea \label{beta}
\beta&=&\sqrt{\int_{\Lambda}^{\infty}dk|(w_{Ik},\phi_B)|^2},\\
\langle \hat{n}_U\rangle &=&\beta^{-2}\int_{\Lambda}^{\infty}\sinh^2{r_{k/a}} |(w_{Ik},\phi_B)|^2dk,\label{n}\\
\beta'&=&\beta^{-1}\int_{\Lambda}^{\infty}(w_{Ik},\phi_B)(w_{Ik},\phi_B^\star)dk,\label{betaprime}
\eea
where $r_{\Omega}=\textrm{arctanh}(e^{-\pi \Omega})$.

A local mode for Bob is specified by introducing an infrared cutoff in the frequencies, selecting only right moving waves and then renormalizing the following Gaussian mode:
\begin{eqnarray}
\label{modeAfunction}
\phi(x,0) &=& \frac{1}{\sqrt{N\sqrt{2\pi}}}\exp\left[-\frac{x^2}{L^2} + i \frac{N}{L}x\right]\nonumber \\ \nonumber \\
\partial_t{\phi}(x,0) &=& -i\frac{N}{L}{\phi}(x,0),
\end{eqnarray}
where $L$ characterizes the width of the mode, $N/L$ is the characteristic frequency of the mode and the cut-off $\Lambda$ is device dependent. Therefore Bob's mode is given by:
\bea
\phi_B(x,t)=|N_{\phi}|\int_{\Lambda}^{\infty}(u_{l},\phi)u_{l}(x,t)dl
\eea
which is a superposition of  strictly positive frequency Minkowski waves. Numerical checks show that this wave packet is Gaussian-like and localized about $x=0$.

We assume that an accelerated observer, Rob, passes through the centre of this mode at $t=0$ therefore we choose the origin of the Rindler frame accordingly. Equivalently we can keep the accelerated frame fixed and translate the mode $\phi_B(x)\rightarrow\phi_B(x-1/a)$. 

Using:
\bea
(u_l,\phi)&=&\frac{N+|l| L}{2\sqrt{|l| N\sqrt{2 \pi}}}e^{-\frac{(l L-N)^2}{4}},\\
(w_{Ik},u_{l})&=& \frac{i}{2\pi} \frac{e^{\frac{\pi k}{2a}}}{\sqrt{|k l|}}\left(\frac{l}{a}\right)^{i k/a}\Gamma(1-\frac{i k}{a}),\\
(w_{Ik},u_{l}^\star)&=&e^{-\pi k/a}(w_{Ik},u_{l}),
\eea
substituted into:
\bea
(w_{Ik},\phi_B(x-\tfrac{1}{a}))&=&|N_{\phi}|\int_{\Lambda}^{\infty}dle^{-\tfrac{il}{a}}(u_l,\phi)(w_{Ik},u_l), \qquad \\
(w_{Ik},\phi_B(x-\tfrac{1}{a})^\star)&=&|N_{\phi}|\int_{\Lambda}^{\infty}dle^{\tfrac{il}{a}}(u_l,\phi)^\star(w_{Ik},u_l^\star), \qquad
\eea
and eventually put into \eqref{beta}-\eqref{betaprime} the quantities $\beta$, $\beta'$ and $\langle \hat{n}_U\rangle$ for the localized mode can be calculated numerically.

In \cite{Dragan2012,Dragan2012b} an explicit detector model was used which was assumed to have finite extent $L$ and perform measurements of the field at $t=0$. The detector cut-off wavelength, $1/\Lambda$, was therefore naturally related to the size of the detector, since wavelengths larger than this would be difficult to resolve. It was further assumed that the whole detector was approximately accelerating at a constant proper acceleration, which placed the constraint $aL\ll1$ on the magnitude of accelerations that could be explored for a given size of the mode. Here go beyond this limit, and in particular investigate large accelerations. To do so we assume that the detector can focus Bob's mode, $\phi_B$, down to a size small enough so that it is measured in a small neighbourhood about the centre of the detector, $\xi=0$ \cite{footmeasuretime}. This position corresponds to the path followed by the hypothetical point-like accelerated observer Rob. Thus, while the focussing lens and other detector components are assumed to be rigid, the measurement itself takes place in a region where the acceleration and proper-time have an approximately unique value.

\emph{--Entanglement.\thinspace}
For two-mode Gaussian states the entanglement as measured by the logarithmic negativity is given by \cite{logneg}:
\bea \label{En}
	E_N=\max\left(0,-\ln \tilde{\nu}_{-}\right),
\eea
where $\tilde{\nu}_{\pm}=\sqrt{(\Delta\pm \sqrt{\Delta^2-4\, \text{det}\,\sigma})/2}$ are the symplectic eigenvalues of the partially transposed covariance matrix and $\Delta =\sigma_{11}\sigma_{22}-\sigma_{12}^2+\sigma_{33}\sigma_{44}-\sigma_{34}^2-2\sigma_{13}\sigma_{24}+2\sigma_{14}\sigma_{23}$.

In the large acceleration limit, the center of Bob's Gaussian mode asymptotes to the position $x=0$ and the penetration of the mode through the horizon stabilizes to an approximately constant value i.e., half of Bob's mode is located outside Rob's horizon. On the other hand the average vacuum particle number continues to grow linearly. This happens because the number of Unruh particles from the vacuum follows a Bose-Einstein distribution with most of the particles in modes below the frequency $k<a$; at low accelerations these modes are below the cutoff and therefore unobservable, but as the acceleration increases more and more particles come into range of the detector. Eventually at infinite acceleration, there will be an infinite number of particles in the mode. The interplay of these two behaviors leads to an interesting conclusion not previously realized in global mode studies of the entanglement degradation in non-inertial frames: the entanglement extractable by Alice and Rob for large enough acceleration will always lead to an entanglement sudden death. To see this we note that the state becomes separable when the smallest symplectic eigenvalue of the partially transposed state, $\tilde{\nu}_-$, is unity \cite{Adesso2004} i.e., when, $\Delta=1+\det \sigma$. For our covariance matrix this implies:
\bea
|\beta|^2=\langle \hat{n}_U\rangle\left(1-\frac{ |\beta'|^2}{1+\langle \hat{n}_U\rangle}\right).
\eea
In the considered range of parameters $|\beta'|^2$ is small and so the entanglement sudden death occurs, for all values of initial squeezing, when the average vacuum particle number is approximately as large as $|\beta|^2$.
\begin{figure}[h]
       \centering
        \includegraphics[width=0.5\textwidth]{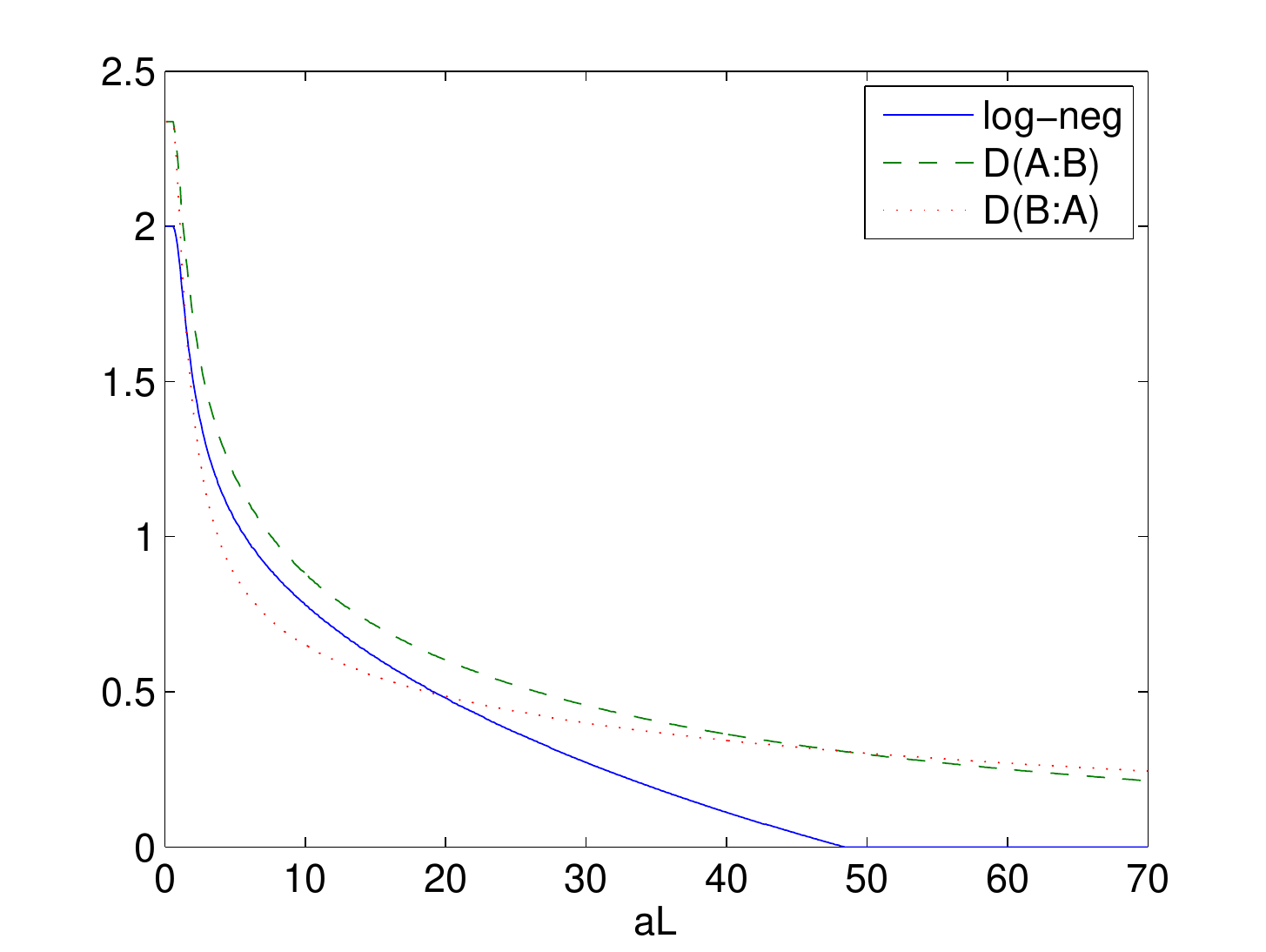}
	\caption{(Color
  online) A plot of logarithmic negativity \(E_N\) (blue, solid), discord \(D(A:B)\) (green, dashed) and discord \(D(B:A)\) (red, dotted) as functions of acceleration, where the state being considered is the two-mode squeezed state, with squeezing parameter \(s=1\), and with Bob's mode in the localized Gaussian, $\phi_B(x-\tfrac{1}{a})$ with $N=6$ and cut-off value chosen to $\Lambda L= \frac{1}{3}$. We see that, unlike when Unruh modes are used, the entanglement experiences sudden death at a finite acceleration. The quantum discord, however, remains even beyond this point of entanglement extinction.}
        \label{localplot1}
\end{figure}

In Fig.~\ref{localplot1} we compute the logarithmic negativity for the localized state as a function of acceleration up to \(a L=70\). The results of this calculation confirm that the entanglement in this state experiences sudden death at a finite acceleration. This is in stark contrast to the entanglement behavior when considering Unruh modes, which we will now elaborate.

Typically, in the literature on entanglement degradation in non-inertial frames, either plane wave modes \cite{Alsing2003, Fuentes2005} are used in which the inaccurate single-mode approximation is applied or Unruh modes are used \cite{Bruschi2010} in which formally the same Bogolyubov transformation occurs but the approximation doesn't have to be invoked. However in the latter approach the analyzed entanglement degrades as a function of the frequency of the considered Unruh mode and cannot be directly interpreted as entanglement degradation due to the acceleration of the observer, which we will now show.

A typical example of an Unruh mode $u_{\Omega}(x,t)$ studied in the literature is:
\bea\label{Unruhmodedefinition}
u_{\Omega}\sim\left(\frac{x-t}{l_\Omega}\right)^{i\Omega},
\eea
with the modes parameterized by a dimensionless index $\Omega$ and where $l_\Omega$ is an arbitrary $\Omega$-dependent constant that has units of length and defines a global phase of the mode. In previous studies \cite{Bruschi2010} the state typically considered was $(1+\hat{a}^{\dagger}\otimes \hat{b}^{\dagger}_\Omega)|0\rangle_M$ where $\hat{b}_\Omega$ is the annihilation operator corresponding to the Unruh mode and the entanglement observed by Alice and Rob was calculated as a function of $\Omega$. Since Unruh modes are orthogonal $(u_\Omega, u_{\Omega'})=\delta(\Omega-\Omega')$ the initial state was being changed to an orthogonal one as $\Omega$ was being varied.

However it is possible to fix the state with a given index $\Omega_0$ and study how much entanglement can be extracted by Alice and Rob from that state as a function of Rob's acceleration. For any acceleration $a$ we calculate Rob's optimal mode using \eqref{RobOptDet}. We find that the result is just the Rindler mode:
\beq \label{UnruhOptDetector}
\psi_R=w_{Ik},
\eeq
where Rob has to tune the frequency of that mode to $k=a \Omega_0$. The fact that the optimal mode turns out to be global is an inevitable consequence of using global, unphysical Unruh modes of the state. We therefore see that in the limit of an infinitely sized detector the optimial detection model introduced in \cite{Dragan2012} becomes equivalent to the preparation and measurement schemes that were previously employed to study entanglement degradation in the case of Unruh mode entanglement \cite{Bruschi2010}. 

Moreover we find that the overlap between \eqref{Unruhmodedefinition} and \eqref{UnruhOptDetector} does not depend on acceleration. Since any measure of entanglement depends only on this overlap, the maximal entanglement accessible by Alice and Rob does not depend on Rob's acceleration:
\bea
E_N(a)=\text{const}_{\Omega_0}, \quad a>0.
\eea

The same thing happens when Alice and Bob prepare a two-mode squeezed state $\exp{\left[s(\hat{a}^{\dagger}\hat{b}^{\dagger}_U+\hat{a}\hat{b}_U)\right]}|0\rangle_M$. In this case, we can explicitly calculate all the parameters (\ref{beta})-(\ref{betaprime}) that go into formula for the entanglement (\ref{En}) and find:
\bea\label{Unruhparams1}
\beta&=&\cosh r_{\Omega_0},\\
\langle \hat{n}_U\rangle&=&\sinh^2r_{\Omega_0},\label{Unruhparams2}\\
\beta'&=&0.\label{Unruhparams3}
\eea
Equations (\ref{Unruhparams1})-(\ref{Unruhparams3}) are only dependent on the fixed, dimensionless parameter $\Omega_0$, therefore the acceleration dependence results in a flat line for the entanglement profile.

The above results are in a stark contrast with all the previous results on entanglement degradation due to acceleration. We conclude that the above-mentioned dependencies are observed only because the entangled states of Unruh modes under consideration were not fixed but varied with $\Omega$.


\emph{--Gaussian quantum discord. \thinspace}
We now investigate the non-classical correlations using quantum discord. The Gaussian quantum discord, involving optimization over Gaussian measurements on subsystem \(B\), was found in \cite{Adesso2010} to be:
\begin{equation}\label{discord}
D(A:B) = f(\sqrt{B})-f(\nu_-) - f(\nu_+) +f(\sqrt{E}),
\end{equation}
where \(\nu_{\pm}\) are the (non-partially transposed) symplectic eigenvalues of the covariance matrix (\ref{covariance}), $f(x) = \left(\frac{x+1}{2}\right) \log\left[\frac{x+1}{2}\right] -\left(\frac{x-1}{2}\right) \log\left[\frac{x-1}{2}\right]$ and:
\begin{widetext}
\begin{eqnarray}
E&=& \left\{  
\begin{array}{l}
\frac{{2 C^2+\left(-1+B\right) \left(-A+\, \text{det}\,\sigma\right)+2 |C| \sqrt{C^2+\left(-1+B\right) \left(-A+\, \text{det}\,\sigma\right)}}}{{\left(-1+B\right){}^2}} ~~~\text{for}\quad\left (\, \text{det}\,\sigma-A B  \right) {}^2 \le \left (1 +   B \right) C^2 \left (A + \, \text{det}\,\sigma \right);\\ \\
\frac{{A B-C^2+\, \text{det}\,\sigma-\sqrt{C^4+\left(-A B+\, \text{det}\,\sigma\right){}^2-2 C^2 \left(A B+\, \text{det}\,\sigma\right)}}}{{2 B}}\quad \hbox{otherwise} ,
\end{array} \right.
   \nonumber
\end{eqnarray}
\end{widetext}
where $A=\sigma_{11} \sigma_{22}-\sigma_{12}\sigma_{21}$; $ B=\sigma_{33} \sigma_{44}-\sigma_{34}\sigma_{43}$; $C=\sigma_{13} \sigma_{24}-\sigma_{14}\sigma_{23}$. The physical meaning of these terms is as follows: \(f(\sqrt{B})\) is the von Neumann entropy of the reduced state of \(B\), \(f(\nu_-) + f(\nu_+)\) is the entropy of the global system, and \(f(\sqrt{E})\) is the entropy of the subsystem \(A\) after a Gaussian measurement has been performed on \(B\), where the measurement is chosen to minimize this quantity (i.e. learn the most possible about \(A\)). 

It is worth noting that discord is not symmetric between \(A\) and \(B\); in general \(D(A:B) \neq D(B:A)\) and one can find states that have zero quantum discord in one direction but not in the other.

Note that  while there is circumstantial evidence that Gaussian measurements are optimal for Gaussian states \cite{Giorda2012}, there is no proof of this in general and so it is possible that the Gaussian discord generally overestimates the true value of discord. Using equation (\ref{discord}) we investigate the quantum discord of localized Gaussian modes for larger accelerations. We computed that in the \(aL \rightarrow \infty\) limit the discord vanishes either when the optimized measurement is over Alice’s subsystem or when the optimized measurement is over Rob’s subsystem. This is different from the results found previously \cite{Datta2009}. 

We note that in the original paper by Datta \cite{Datta2009} the single mode approximation was invoked and superpositions of Fock states were used as the initial state.  Since it was shown in \cite{Bruschi2010} that the single mode approximation is inaccurate, we will reinterpret Datta's results by replacing the plane wave mode with that of an Unruh mode. In this case the degradation effects are parameterized by the frequency of the initial Unruh mode rather than the acceleration. We will now investigate the discord for Unruh modes in the two-mode squeezed state as a function of this initial Unruh mode frequency. 

\begin{figure}[h]
       \centering
        \includegraphics[width=0.5\textwidth]{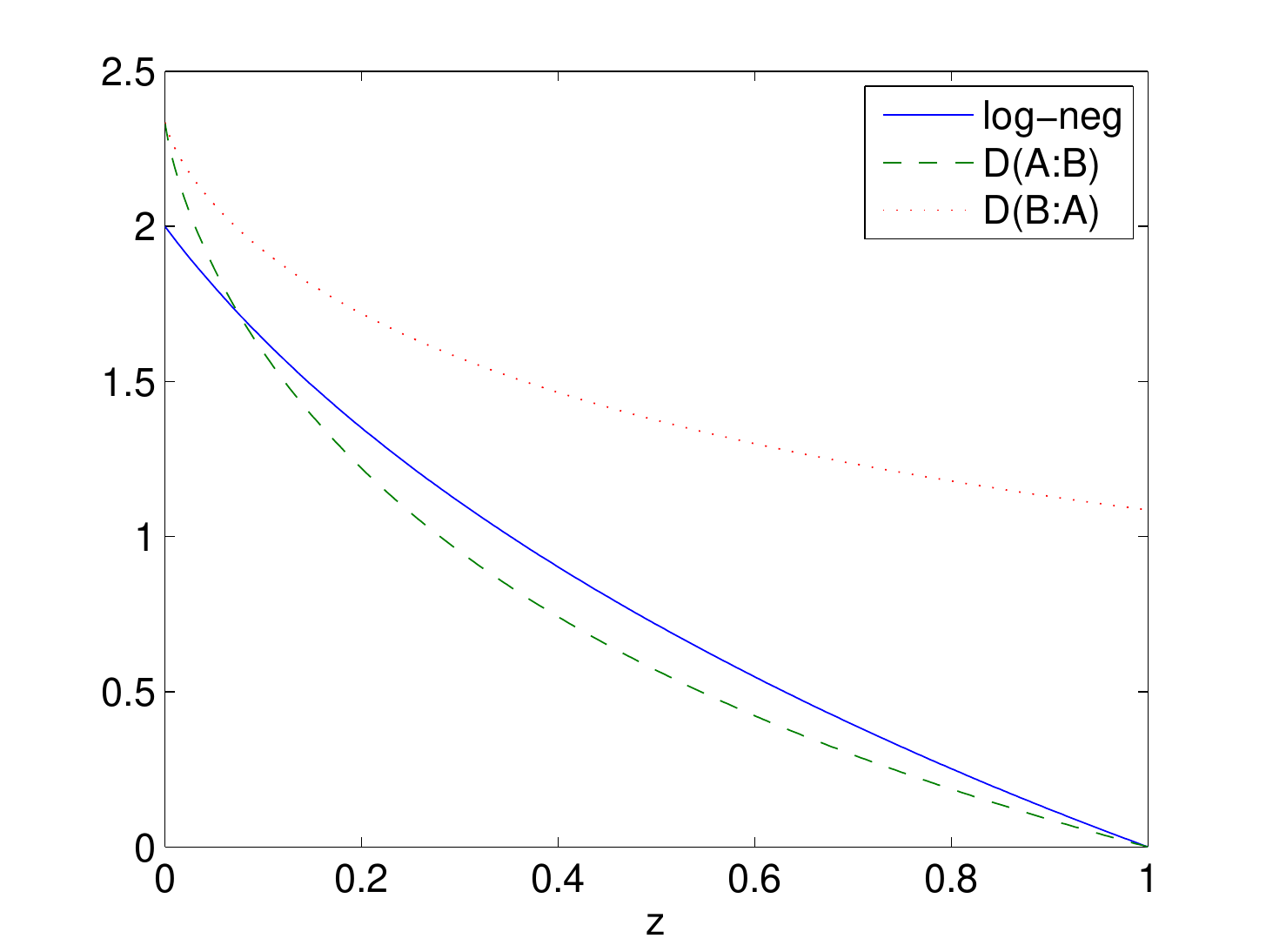}
	\caption{(Color
  online) A plot of logarithmic negativity \(E_N\) (blue, solid), discord \(D(A:B)\) (green, dashed) and discord \(D(B:A)\) (red, dotted) as functions of \(z=e^{-2 \pi \Omega}\), where the state in consideration is a two-mode squeezed state using Unruh modes with squeezing parameter \(s=1\). We see that \(E_N\) and \(D(A:B)\) decay to zero as \(\Omega \rightarrow 0\) whereas \(D(B:A)\) asymptotes to a finite value.\label{unruhplot}}
    \end{figure}

Fig.~\ref{unruhplot} shows the quantum discord (both \(D(A:B)\) and \(D(B:A)\)) as a function of \(z \equiv e^{-2 \pi \Omega}=\langle \hat{n}_U \rangle/ (\langle \hat{n}_U \rangle+1)\) where the function $z$ has been chosen to rescale the entire frequency domain down to the unit interval. These were computed using the parameters (\ref{Unruhparams1})-(\ref{Unruhparams3}) input in the covariance matrix of \cite{Dragan2012b} and using the Gaussian discord as given by \cite{Adesso2010}. We see that as $\Omega\rightarrow 0$ ($z=1$) the discord \(D(B:A)\), asymptotes to a finite value. These results match the behavior previously observed using the Unruh-mode Fock states \cite{Datta2009}.  We also find an analytic expression for this residual discord as a function of the squeezing parameter: $\lim_{\Omega\rightarrow0}D(B:A)=2\log (\coth s)\sinh^2 s.$ Curiously, this function quickly asymptotes to a value of \(1/\ln2\) for large squeezing values \(s\). This means that even though the correlations could be arbitrarily large in the inertial frame (arbitrarily large squeezing), in the accelerating frame they will always be bounded.  On the other hand, we observe that when the optimized measurement is over Rob's subsystem, the discord \(D(A:B)\) decays to zero in the $\Omega\rightarrow 0$ limit. It is worth pointing out that this novel result is computationally intractable in the analogous Fock state scenario.  

After the first version of this paper appeared the results of \cite{Gerardo2012} were brought to our attention. In this recent work the discord was calculated for a two-mode squeezed state in the Unruh mode setting. We have verified that using an Unruh mode initial state and the optimal detector mode (\ref{UnruhOptDetector}) in our setting, the covariance matrix (\ref{covariance}) becomes identical to the one found in \cite{Gerardo2012} after accounting for the different squeezing definition $s\rightarrow-s$ and making the replacement $r=\text{Arcsinh}(\sqrt{\langle \hat{n}_U\rangle})$ which follows from equation (\ref{Unruhparams2}). However, it should be noted that the discord calculated in \cite{Gerardo2012} was a Renyi entropy characterized by $\alpha=2$, whereas $\alpha=1$ corresponds to the usual von Neumann entropy that we have used here. For this reason we obtain different numerical values for the discord. Nevertheless the same qualitative features are present.

Somewhat surprisingly we have shown that in the $\Omega\rightarrow0$ limit some correlations between Alice's and Rob's measurements remain even though the average vacuum particle number (i.e., noise contaminating Rob's measurements) becomes infinite. This occurs because of the relationship between $\beta$ and $\langle \hat{n}_U \rangle$ for Unruh modes (\ref{Unruhparams1})-(\ref{Unruhparams2}), which in the limit $\Omega\rightarrow0$, becomes \(\beta \approx \sqrt{\langle \hat{n}_U \rangle}\). Thus the off-block-diagonal terms (i.e. those that represent correlations) in the covariance matrix (\ref{covariance}) increase as \(\sqrt{\langle \hat{n}_U \rangle}\) comparably with some of the diagonal terms. 

There is another instance of when the average vacuum particle number becomes infinite in a completely different setup. This happens in the localized Gaussian scenario in the limit of infinite acceleration. However, in this case the correlation entries remain small (compared with the average vacuum particle number) and therefore no discord remains.

The realization that there is a qualitative difference between the localized and delocalized settings is important because of the considerable amount of work in the literature that uses delocalized Unruh modes in order to study acceleration-induced entanglement degradation. The use of these modes is for computational ease and it has been argued that they should produce qualitatively the same behavior as would more realistic, localized setups. We have demonstrated here that this is in fact not the case. 
\emph{--Conclusions.\thinspace}
We have used the formalism developed and explored in \cite{Dragan2012, Dragan2012b} to study the Unruh-degradation of quantum correlations in two-mode squeezed states, and in particular to understand the difference between the cases of localized Gaussian modes and the delocalized Unruh modes so often used in the literature \cite{Datta2009,Brown2012,biglist}. Although most past studies that used Unruh modes did so with Fock states, rather than squeezed states, we have shown that the degradation of quantum correlations are qualitatively equivalent between the two cases. In regards to this we have found that the non-vanishing quantum discord previously observed in the case of a zero frequency Unruh mode, appears only to be true when the optimized measurement is over Alice's subsystem but not so when it is over Rob's.

Comparing the cases of localized and delocalized modes reveals qualitatively different results, indicating that many of the conclusions presented in previous literature \cite{biglist} may have to be reconsidered. In particular we have found in the localized modes that the entanglement reaches a point of sudden death at a finite acceleration, and that the quantum discord vanishes in the infinite acceleration limit (both \(D(A:B)\) and \(D(B:A)\)). 

\emph{--Acknowledgments.\thinspace}
This work was supported in part by the Natural Sciences and Engineering Research Council of Canada.


\begin{thebibliography}{99}
\bibitem{footnoteCutoff}
We are compelled to introduce an IR cut-off because of divergences in the low Rindler frequency limit of the integrand. However this cut-off is also required on physical grounds since no detector can resolve arbitrary low frequencies. Indeed the wavelength of the IR cut-off is naturally related to the physical dimensions, $L$, of the detector. One might also have expected a UV cut-off. However, for the Gaussian mode considered the integrand vanishes for large $k$, and therefore the integrals are independent of any (sufficiently) large UV cut-off.
\bibitem{footmeasuretime}
Equivalently, the detection could occur in a neighbourhood of $\xi=0$ without focusing the mode if one allowed for a measurement time $L/c$.
\end{thebibliography}
\end{document}